\documentclass[prl,twocolumn,preprintnumbers,amsmath,amssymb]{revtex4}
\usepackage{textcomp}
\usepackage{amssymb}
\usepackage{amsmath}
\usepackage{amsthm}

\usepackage{graphicx}% Include figure files
\usepackage{txfonts}
\begin{document}

\preprint{}
\title{Realization of Tunable Photonic Spin Hall Effect by Tailoring the Pancharatnam-Berry Phase}% Force line breaks with \\
\author{Xiaohui Ling$^{1,3}$}
\author{Xinxing Zhou$^2$}
\author{Weixing Shu$^2$}
\author{Hailu Luo$^{1,2}$}\email{hailuluo@hnu.edu.cn}
\author{Shuangchun Wen$^{1,2}$}\email{scwen@hnu.edu.cn}

\affiliation{$^1$Key Laboratory of Optoelectronic Devices and
Systems of Ministry of Education and Guangdong Province,
College of Optoelectronic Engineering, Shenzhen University, Shenzhen 518060\\
$^2$Key Laboratory for Micro-/Nano-optoelectronic Devices of
Ministry of Education, College of Physics and Microelectronic
Science, Hunan University, Changsha
410082, China\\
$^3$Department of Physics and Electronic Information Science,
Hengyang Normal University, Hengyang 421002, China}

\date{\today}% It is always \today, today,
\maketitle

\textbf{Recent developments in the field of photonic spin Hall
effect (SHE) offer new opportunities for advantageous measurement of
the optical parameters (refractive index, thickness, etc.) of
nanostructures and enable spin-based photonics applications in the
future. However, it remains a challenge to develop a tunable
photonic SHE with any desired spin-dependent splitting for
generation and manipulation of spin-polarized photons. Here, we
demonstrate experimentally a scheme to realize the photonic SHE
tunably by tailoring the space-variant Pancharatnam-Berry phase
(PBP). It is shown that light beams whose polarization with a
tunable spatial inhomogeneity can contribute to steering the
space-variant PBP which creates a spin-dependent geometric phase
gradient, thereby possibly realizing a tunable photonic SHE with any
desired spin-dependent splitting. Our scheme provides a convenient
method to manipulate the spin photon. The results can be
extrapolated to other physical system with similar topological
origins.}

Spin Hall effect (SHE) is a transport phenomenon in which an
electric field applied to spin particles results in a spin-dependent
shift perpendicular to the electric field
direction~\cite{Murakami2003,Sinova2004,Wunderlich2005}. It plays an
important role in the spintronics since it can provide an effective
way for generation, manipulation, and detection of spin-polarized
electron which define the main challenges in spintronics
applications~\cite{Wolf2001,Awschalom2007,Chappert2007}. Photonic
SHE manifests itself as the spin-dependent splitting of
light~\cite{Onoda2004,Bliokh2006,Hosten2008}, which can be regarded
as a direct optical analogy of the SHE in electronic systems.
Recently, photonic SHE has been proved to be an advantageous
metrological tool for characterizing the parameters of
nanostructures~\cite{Zhou2012a,Zhou2012b}. More importantly, it
offers new opportunities for manipulating spin photons and
developing next generation of spin-controlled photonic devices as
counterparts of recently presented spintronics
devices~\cite{Wolf2001,Shitrit2013}. Similar to those in the
spintronics devices, one major issue in spin-based photonics is to
develop a tunable photonic SHE with any desired spin-dependent
splitting for generation, manipulation, and detection of
spin-polarized photons.

Photonic SHE is generally believed to be a result of topological
spin-orbit interaction which describes the coupling between the spin
(polarization) and the trajectory (orbit angular momentum) of light
beam propagation. The spin-orbit interaction corresponds to two
types of geometric phases: the Rytov-Vladimirskii-Berry phase
associated with the evolution of the propagation direction of
light~\cite{Onoda2004,Bliokh2006,Hosten2008} and the
Pancharatnam-Berry phase (PBP) related to the manipulation with the
polarization state of light~\cite{Bliokh2008a,Bliokh2008b}. The
spin-orbit interaction associated with the Rytov-Vladimirskii-Berry
phase is exceedingly small, and the detection of the corresponding
photonic SHE needs multiple reflections~\cite{Bliokh2008b} or weak
measurements~\cite{Hosten2008,Luo2011}. Recently, the strong
spin-orbit interaction related to the PBP in metasurface, a
two-dimensional electromagnetic nanostructure, enables one to
observe a giant photonic SHE~\cite{Shitrit2011,Yin2013,Li2013}.
However, the PBP in the metasurface is fixed and unadjustable, once
the nanostructure is manufactured. Despite of recent experimental
efforts, it remains a challenge to realize a photonic SHE with a
high tunability.

In this work, we report an experimental realization of tunable
photonic SHE via tailoring the space-variant PBP contributed by a
light beam with transversely inhomogeneous polarization. We show
that the space-variant polarization is able to invoke the spin-orbit
interaction even when propagating through a homogeneous waveplate
and produces the space-variant PBP as the inhomogeneous metasurface
does, with a high tunability the latter cannot achieve. As the beam
with any desired space-variant polarization can be easily generated
by many methods (see Ref.~\cite{Zhan2009} for a review and the
references therein), involving the contribution of the incident
polarization distribution to the space-variant PBP will introduce a
convenient degree of freedom to manipulate the photonic SHE.
\\

\begin{figure}
\includegraphics[width=8cm]{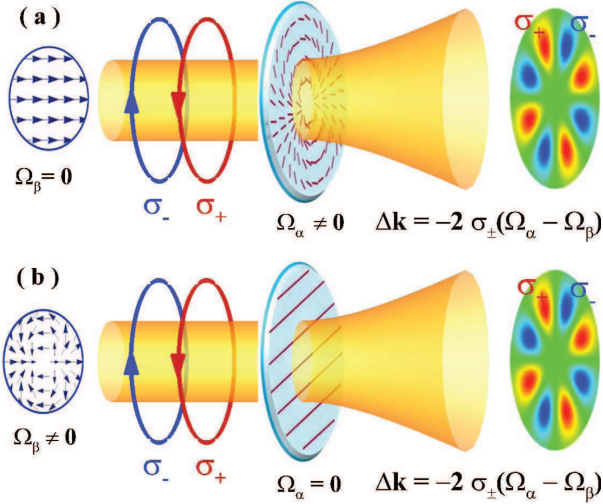}
\caption{\label{Fig1} Schematic illustration of the photonic SHE
with multi-lobe, spin-dependent momentum splitting. (a) A collimated
Gaussian beam with homogeneous polarization (spatial rotation rate
$\Omega_\beta=0$) propagates through an anisotropic metasurface with
spatial inhomogeneity ($\Omega_\alpha\neq0$). The geometric phase
gradient along the azimuthal direction refracts light in twisted
directions and results in the spin-dependent momentum splitting. (b)
A beam with tunable spatial rotation rate ($\Omega_\beta\neq0$)
passes through a homogeneous anisotropic waveplate
($\Omega_\alpha=0$), which can also induce a geometric phase
gradient and momentum splitting.}
\end{figure}

\noindent\textbf{RESULTS}\\
\noindent\textbf{Theoretical analysis}. For an inhomogeneous
waveplate or metasurface, the space-variant optical axis direction
produces a spatial rotation rate and applies a space-variant PBP to
a light beam that passes through it, due to the spin-orbit
interaction~\cite{Shitrit2013,Bliokh2008a}. This inhomogeneous
geometric phase forms a spin-dependent geometric phase gradient and
generates a spin-dependent shift ($\Delta k$) in momentum space
[Fig.~\ref{Fig1}(a)]. Indeed, the inhomogeneous birefringent
waveplate is not the sole way to produce the PBP. Providing the
incident beam takes a spatial rotation rate, namely exhibits an
inhomogeneous polarization, the PBP and spin-dependent momentum
splitting should still occur [Fig.~\ref{Fig1}(b)]. Recent works
about the spin-dependent splitting occurring at
metasurfaces~\cite{Shitrit2011,Yin2013,Li2013} can be seen as
special cases where only the waveplate is spatially inhomogeneous.
The scheme proposed here show that a beam with tunable inhomogeneous
polarization can also generate the spin-dependent momentum shift
even though it passes through a homogenous waveplate.

\begin{figure}
\includegraphics[width=8.5cm]{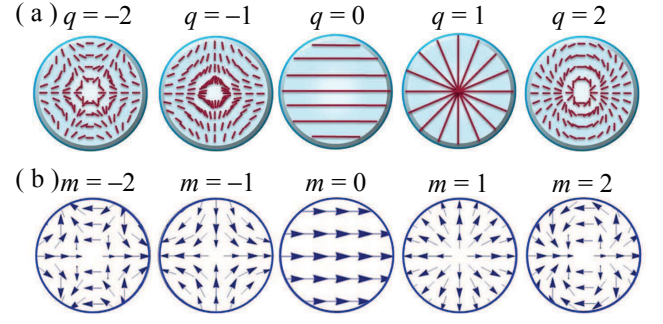}
\caption{\label{Fig2} (a) Schematic examples of the waveplate
geometries for different spatial inhomogeneity. The tangent to the
lines shown indicates the direction of local optical axis. (b)
Schematic examples of the polarization of CVBs for different spatial
inhomogeneity. The arrows represent the local polarization
directions.}
\end{figure}

We now develop a unified theoretical model of the photonic SHE for
either the polarization of light beam or the waveplate exhibits a
spatial inhomogeneity. First, we consider that only the waveplate is
inhomogeneous. It is assumed that the waveplate is a uniaxial
crystal with its optical axis direction specified by a space-variant
angle $\alpha(r,\varphi)=q\varphi+\alpha_0$, where $q$ is a
topological charge, $\varphi=\arctan(y/x)$ the local azimuthal
angle, and $\alpha_0$ the initial angle of local optical axis
direction~\cite{Marrucci2006}. Some examples of the inhomogeneous
waveplate geometries are shown in Fig.~\ref{Fig2}(a). It is the
spatial version of a rotationally homogeneous
waveplate~\cite{Bliokh2008a}. Similar to the temporal rotation rate,
the spatial rotation rate of the inhomogeneous waveplate in the
polar coordinate is
$\Omega_\alpha=\text{d}\alpha(r,\varphi)/\text{d}\xi$, where
$\text{d}\xi=r\text{d}\varphi$ is the rotating route with $r$ the
radius of the waveplate.

Next, we consider only the polarization of the incident beam
exhibits a spatial inhomogeneity. A linear polarization, with its
electric field described by a Jones vector, is given as
\begin{eqnarray}
E_{\text{in}}(x,y)=\left(\begin{array}{ccc}
 \cos\beta\\
 \sin\beta\end{array}\right)E_0(x,y),\label{linear}
\end{eqnarray}
where $\beta$ is the polarization angle and $E_0(x,y)$ the complex
amplitude. Generally, $\beta$ is a constant, and Eq.~(\ref{linear})
represents homogeneous linear polarizations with $\beta=0$ and
$\pi/2$ respectively indicating linear $x$- and $y$-polarizations.

If $\beta$ is a position-dependent function, e.g.,
$\beta(r,\varphi)=m\varphi+\beta_0$, Eq.~(\ref{linear}) represents a
beam with its polarization having cylindrical symmetry, i.e., the
cylindrical vector beam (CVB). Here, $m$ is also a topological
charge and $\beta_0$ the initial polarization angle for $\varphi=0$.
Similar to the inhomogeneous anisotropic waveplate, the
inhomogeneous linear polarization has a spatial rotation rate
$\Omega_\beta=\text{d}\beta(r,\varphi)/\text{d}\xi$ which can be
tailored by suitably designing the polarization distribution of the
CVBs [see Fig.~\ref{Fig2}(b) for examples].

In the most general case, both the polarization of light and the
waveplate possess spatial rotation rates. Their relative rotation
rate can be regarded as $\Omega_\gamma=\Omega_\alpha-\Omega_\beta$.
Analogous to the rotational Doppler effect, the PBP can be written
as~\cite{Bliokh2008a}
\begin{eqnarray}
\Phi_{\text{PB}}=-2\sigma_{\pm}\int
\Omega_\gamma\text{d}\xi=-2\sigma_{\pm}[(q-m)\varphi+(\alpha_0-\beta_0)],\label{pbp}
\end{eqnarray}
where $\sigma_+=+1$ and $\sigma_-=-1$ representing the left and
right circular polarization photons, respectively. Note that the PBP
has a spin-dependent spiral structure which is induced by the
topological spin-orbit
interaction~\cite{Shitrit2011,Niv2008,Ling2012}. The factor $-2$
arises because the spin angular momentum vector of photon is
reversed, leading to the conversion of spin angular momentum to its
orbital parts twice the magnitude of the photon¡¯s spin angular
momentum~\cite{Garetz1981,Dahan2010}. This indicates that the PBP
for beams evolving in a system with nonzero spatial rotation rate
would cause a spin-dependent splitting in $k$ (momentum) space,
i.e., a manifestation of the photonic SHE.

\begin{figure}
\includegraphics[width=8cm]{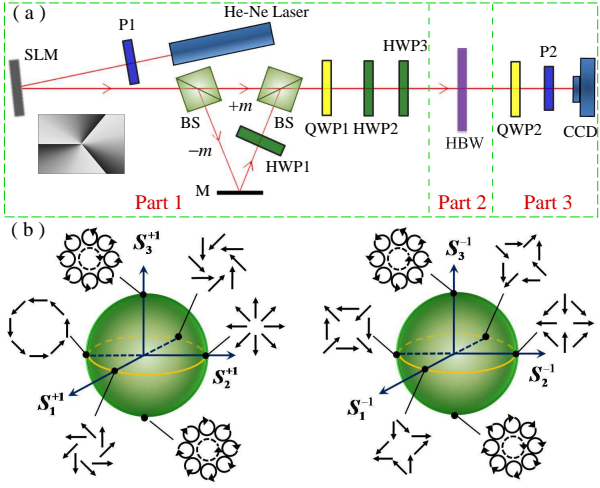}
\caption{\label{Fig3} (a) Experimental apparatus for generating the
CVBs and measuring the tunable photonic SHE. Part 1: A collimated
and horizontal polarized $\text{TEM}_{00}$ beam from a He-Ne laser
(632.8 nm, 21mW) followed by a polarizer (P1) is converted into a
vortex-bearing $\text{LG}_{0,m}$ beam using a reflective phase-only
spatial light modulator (SLM) (Holoeye Pluto-Vis, Germany). Using a
modified Mach-Zender interferometer comprised of two beam splitters
(BS) and one mirror (M), a $\text{LG}_{0,-m}$ beam is produced by an
odd number of reflections. A half-wave plate (HWP1) with $45^\circ$
inclined to horizontal direction converts the horizontal
polarization to vertical one and a quarter-wave plate (QWP1) gives
the two beams orthogonal circular polarizations. Their collinear
superposition generates a CVB on the equator of the higher-order
Poincar\'{e} sphere. In order to determine the concrete point on the
equator, two cascaded half-wave plates (HWP2 and HWP3) are
employed~\cite{Zhan2009,Supplementary}. Part 2: The CVB passes
through the homogeneous birefringent waveplate (HBW). The photonic
SHE can be detected by Part 3 (a quarter-wave plate followed by a
polarizer and a CCD camera) which is a typical setup for measuring
the Stokes parameters $S_3$. The inset: An example of the phase
picture displayed on SLM. (b) Higher-order Poincar\'{e} spheres.
Left panel for the case of $m=1$ and the right panel for $m=-1$. The
equator represents the linearly polarized CVBs, the poles circularly
polarized vortex beams, and intermediate points between the poles
and equator elliptically polarized CVBs. The solid circles with
arrows in the north/south poles indicate locally circular
polarizations, while the dashed ones represent the direction of
vortex.}
\end{figure}

Unambiguously, the PBP is locally varying, and creates a geometric
phase gradient which can be tunable by tailoring the waveplate
geometry and polarization distribution of the CVB. This will lead to
a spin-dependent shift in $k$ space. Under the normal incidence, the
spin-dependent shift $\Delta k=\Delta k_{\text{r}}+\Delta
k_{\varphi}$ can be calculated as the gradient of the PBP:
\begin{eqnarray}
\Delta k=\nabla\Phi_{\text{PB}}
=-2\sigma_\pm(q-m)\hat{e}_{\varphi},\label{kshift}
\end{eqnarray}
where $\hat{e}_\text{r}$ ($\hat{e}_\varphi$) is the unit vector in
the radial (azimuthal) direction. Here $\Delta k$ only has the
azimuthal component $\hat{e}_{\varphi}$, which means that the
spin-dependent splitting occurs also in this direction. Note that
the spin-dependent splitting will vanish if the beam polarization
and inhomogeneous waveplate have the same spatial rotation rate
(i.e., $q=m$). As the polarization rotation rate of the CVB can be
tailored arbitrarily via suitably designing the polarization
distribution of the CVB, it can serve as an independent and
convenient degree of freedom for manipulating the PBP and photonic
SHE.
\\

\noindent\textbf{Experimental results.} To demonstrate the role of
the inhomogeneous polarization of incident beam in the tunable
photonic SHE, a CVB with its polarization exhibiting tunable spatial
rotation rate is generated experimentally and passes through a
homogeneous waveplate [Fig.~\ref{Fig3}(a)]. It is a typical example
of the Fig.~\ref{Fig1}(b). Part 1 of the experiment setup can
produce a CVB on the equator of the higher-order Poincar\'{e} sphere
where the polarization distribution possesses a cylindrical
symmetry~\cite{Holleczek2011,Milione2011} [Fig.~\ref{Fig3}(b)]. The
CVB on the equator is generated by interfering the two beams from a
modified Mach-Zender interferometer. Two cascaded half-wave plates
(HWP2 and HWP3) are employed to determine the polarization
distribution of the linearly polarized CVB. By rotating the optical
axis directions of the two half-wave plates, we can obtain any
desired polarization on the equator of the higher-order Ponicar\'{e}
spheres~\cite{Zhan2009,Supplementary}. The CVB passes through a
homogeneous birefringent waveplate (part 2, HBW) and the Stokes
parameter $S_3$ is measured by the part 3. Modulating the phase
picture displayed on the SLM, we can acquire a CVB with any desired
polarization distribution, and thereby introduce a tunable geometric
phase gradient even if the light beam propagates through a
homogeneous anisotropic waveplate.

\begin{figure}
\includegraphics[width=8cm]{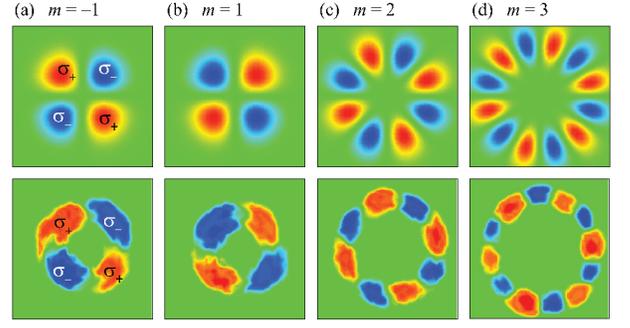}
\caption{\label{Fig4} Tunable photonic SHE for the CVBs exhibiting
different polarization inhomogeneity $m$ under the condition of
$q=0$ and $\alpha_0=0$. The upper row: theoretical calculations; the
lower row: experimental results. Here, $\beta_0=0$.}
\end{figure}

The spin-orbit interaction between the CVB and the HBW brings about
a tailorable PBP and creates a geometric phase gradient in the
azimuthal direction $\hat{e}_{\varphi}$ [see Eqs.~(\ref{pbp}) and
(\ref{kshift})]. This phase gradient can result in a $k$-space
spin-dependent splitting also in this direction. And then the
real-space splitting manifested as multi-lobe patterns can be
directly measured in the far field (Fig.~\ref{Fig4}), which is
attributed to the coherent superposition contributed by the local
$k$-space shift in the near field. For $m=\pm 1$, both the splitting
patterns have four lobes with 2-fold rotational ($C_2$) symmetry.
The only difference is their anti-phase distribution of the $S_3$
pattern due to their just opposite PBP. The splitting pattern for
$m=2$ and 3 respectively show 4-fold ($C_4$) and 6-fold ($C_6$)
rotational symmetry. Actually, the lobe number is equal to $4|m|$.
As $\varphi$ is the azimuthal angle taking values in the range of
$0$ to $2\pi$, $\Phi_{\text{PB}}$ falls in the range of $0$ to
$4m\pi$ for $\sigma_+$ and $0$ to $-4m\pi$ for $\sigma_-$,
respectively. In the spin-dependent splitting pattern, $\sigma_+$
and $\sigma_-$ just has a phase difference of $\pi$, and they
respectively repeat again for each $2\pi$ of $\Phi_{\text{PB}}$.
Therefore, the spin-dependent splitting manifests as a pattern of
alternative $\sigma_+$ and $\sigma_-$ with rotational symmetry, and
the total number of the lobes is $4|m|$, which can be switched by
modulating the topological charge ($m$) of the CVB, i.e., changing
the phase picture shown on the SLM. Note that if a spin-dependent
splitting in the radial direction $\hat{e}_\text{r}$ is desired, a
CVB with polarization inhomogeneity (and the corresponding PBP
gradient) in this direction must be created.

\begin{figure}
\includegraphics[width=8cm]{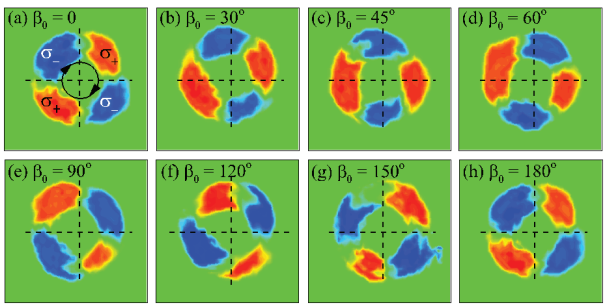}
\caption{\label{Fig5} Revolving the spin-dependent splitting of the
photonic SHE by evolving the polarization of incident CVB along the
equator of higher-order Poincar\'{e} sphere ($m=1$). The arrows in
(a) show the rotation direction.}
\end{figure}

Now let the beam polarization evolve along the equator of the
higher-order Poincar\'{e} sphere where the polarization has the same
spatial rotation rate but different initial polarization angles.
When the beams propagating through a homogeneous anisotropic
waveplate, the PBP could result in a photonic SHE with rotatable
splitting patterns. Figure~\ref{Fig5} shows the experimental results
of the spin-dependent splitting under the evolution of polarization
for $m=1$. Due to the homomorphism between the physical SU(2) space
of the light beam and the topological SO(3) space of the
higher-order Poincar\'{e} sphere, the change of beam polarization of
$180^{\circ}$ corresponds to the rotation of $360^{\circ}$ in the
equator of the sphere. The splitting lobes also revolve
$360^{\circ}$ and return to the original state. This, in turn,
indicates the topological characteristic of the photonic SHE. \\

\noindent\textbf{DISCUSSION}\\The inhomogeneous metasurface which
can produce a desired transverse phase gradient has been employed to
shape the wavefront and manipulate the out-of-plane reflection and
refraction of light based on the generalized Snell's
law~\cite{Yu2011}. Therefore, the metasurface can serve as a
powerful tool for designing optical components like the
inhomogeneous waveplates with locally controllable polarization and
geometric phase. The major issues, such as high losses,
cost-ineffective fabrications, and unadjustable structures, hamper
the development of metasurface technology~\cite{Kildishev2013}. But
no such difficulty exists when a beam with inhomogeneous
polarization is employed to generate the geometric phase. It can
produce the same geometric phase gradient as the inhomogeneous
metasurface does, in particular with a high tunability because a
beam with any desired polarization distribution can be conveniently
achieved by modulating the phase picture displayed on the SLM
(Fig.~\ref{Fig3}). So, the CVB offers a convenient degree of freedom
to manipulate the PBP and photonic SHE.

In summary, tunable photonic SHE has been realized experimentally
via tailoring the space-variant PBP produced by a light beam with
transversely inhomogeneous polarization. By suitably designing the
polarization inhomogeneity, a tunable photonic SHE with any desired
spin-dependent splitting (rotatable multi-lobe patterns) is
realizable. Though the incident beam with spatial-variant
polarization was constructed by a particular method in the
experiment, such a scheme has a general meaning in that
inhomogeneous polarization distributions provide a robust method to
modulate the space-variant PBP, thereby a new way to controllably
manipulate the photonic SHE and spin photon. The resulted SHE is
large enough for direct measurements, in contrast with the indirect
technology using the weak measurement and can be viewed as a
photonic version of the famous Stern-Gerlach experiment. Of
particular interest is that our results can be generalized to other
physical system due to the similar topological origins, such as
vortex-bearing electron
beam~\cite{Bliokh2007,Karimi2012}.\\

\noindent\textbf{METHODS}\\
\noindent\textbf{Experimental measurements of the Stokes parameters
$S_3$}. A quarter-wave plate followed by a polarizer and a CCD
camera [see the part 3 of the Fig.~\ref{Fig3}(a)] is a typical setup
to measure the Stokes Parameter $S_3$ pixel by pixel. By suitably
setting the optical axis direction of the quarter-wave plate and the
transmission axis of the polarizer, we can obtain the intensities of
the left- and right-handed circular polarization components,
respectively. And after a series of calculations, we finally can
obtain the $S_3$ for each pixel.\\

\setlength{\parindent}{0pt}
\textbf{Acknowledgements}\\This research
was supported by the National Natural Science Foundation of China
(Grants No. 61025024, No. 11274106, and No. 11347120), the
Scientific Research Fund of Hunan Provincial Education Department of
China (Grant No. 13B003), and the Doctorial Start-up Fund of
Hengyang Normal University (Grant No. 13B42).\\

\end{document}